\documentclass[%
reprint,
amsmath,amssymb,
aps,
prb,
]{revtex4-2}
\usepackage{graphicx}
\usepackage{dcolumn}
\usepackage{bm}
\usepackage{hyperref}
\usepackage{mathtools}
\usepackage{booktabs}
\usepackage{xcolor,soul}
\usepackage{titlesec}

\newcommand{\eendparagraph}{\\}

\begin{document}
	%\preprint{APS/123-QED}
	
	\title{Two-dimensional hydrodynamic viscous electron flow in annular Corbino rings}

	\author{Sujatha Vijayakrishnan$^{1}$, Z. Berkson-Korenberg$^{1}$, J. Mainville$^{1}$, L.~W. Engel$^{2}$, M.~P. Lilly$^{3}$,  K. W. West$^{4}$, L. N. Pfeiffer$^{4}$  and G. Gervais$^{*,1}$ \vspace{6pt}}

	\affiliation{$^{1}$Department of Physics, McGill University, Montr\'eal,  Qu\'ebec H3A 2T8, CANADA}
	
	\affiliation{$^{2}$National High Magnetic Field Laboratory, Tallahassee, Florida 32310, USA}
	
	\affiliation{$^{3}$Center for Integrated Nanotechnologies, Sandia National Laboratories, Albuquerque, New Mexico 87185, USA}

	\affiliation{$^{4}$Department of Electrical Engineering, Princeton University, Princeton, NJ  08544, USA}
	
	\affiliation{$^{*}$Corresponding author: gervais@physics.mcgill.ca}
	
	\date{\today }
	\begin{abstract}
		
		The concept of fluidic viscosity is ubiquitous in condensed matter systems hosting a continuum where macroscopic  properties can emerge. While an important property of liquids and some solids, only recently the viscosity of electron was shown to play a role in electronic transport experiments. In this letter, we present nonlocal electronic transport measurements in concentric annular rings formed in high-mobility 2DEGs, and the resulting data shows that viscous hydrodynamic flow can occur far away from the source-drain current region. Our conclusion of viscous electronic transport is further corroborated by simulations of the Navier-Stokes equations that are found to be in agreement with our measurements below $T=1~K$. Finally, this work emphasizes the key role played by viscosity via electron-electron ($e-e$) interaction even when the electronic transport is restricted radially, and for which {\it a priori} should  have played no major role.
	\end{abstract}
	
	\maketitle 
	%%%%%%%%%%%%%%%%%%%%%%%%%%%%
	
	{\it Introduction}.---Consider a gas moving through a small constriction: depending on the density of the gas, flow out of the chamber can vary significantly due to the varying mean-free path of its constituent particles. If the mean-free path greatly exceeds the hole size, individual gas particles will diffuse independently with their own trajectories, akin to single particle effusive transport. Conversely, if the mean-free path is much smaller than the hole size, the particles will interact more frequently, ultimately leading to a collective flow guided by the broader dynamics of the ensemble and forming a continuum.  When this is the case, the flow properties of neutral fluids are typically governed by the  Navier-Stokes equations of hydrodynamics, which holds true for the flow of water and other fluids that are ubiquitous in our everyday life.\\
	
	In the case of electrons in metals governed by Fermi liquid theory, somewhat oppositely single particle theoretical descriptions are most often used, neglecting electron viscosity arising from a possible continuum. In narrow graphene constrictions, however, collective effects were shown experimentally to be important under the form of current backflow, forming whirlpools of electrons \cite{Bandurin2016}. 
	Recently, Ahn and Das Sarma \cite{Sarma2022} further proposed that hydrodynamic effects should be observable in the bulk of moderately high electron mobility GaAs/AlGaAs two-dimensional electron gases (2DEGs), provided their scattering parameters place them into a collective regime. In this letter, we present local and nonlocal electronic transport measurements in annular regions formed in GaAs/AlGaAs 2DEGs that have no edges and are therefore entirely bulk. Numerical simulation based on the Navier-Stokes equations for a charged fluid that includes the electron viscosity  proposed by Polini and co-workers \cite{Polini2015}   is found to be in good agreement with experimental results, hence confirming the significance of electron viscosity and viscous drag far away from the source drain current region.\eendparagraph
	
	{\it Lengthscales and hydrodynamic  electron transport }.---In the regime of clean conductors, such as high-mobility 2DEGs, momentum scattering is dominated by electron-electron ($e-e$) interactions, where the mean-free path due to these interactions ($\ell_{MC}$) is much smaller than both the conducting channel width ($W$) as in a Hall bar geometry and the momentum relaxing mean-free path ($\ell_{MR}$) from either impurity or phonon scattering. In the case where $\ell_{MC} \ll W \ll \ell_{MR}$, a collective behaviour of electrons can emerge, that resembles a fluid-like continuum rather than discrete single particle effusive motion. This type of electronic transport is commonly referred to as hydrodynamic transport. This transport regime in metals was first theorized by R. N  Gurzhi \cite{Gurzhi1963, Gurzhi1968} in 1963, and it offers a more complete perspective on electronic transport wherein the resistance to flow is governed by the fluidic viscosity of the electron stream and the geometric characteristics of the conducting channel. This has recently led to a renewed interest in the study of hydrodynamics in electronic systems, and blossomed into a series of theoretical  \cite{Polini2015,Sarma2022,Polini2016,Moore2017,Tobias2019,Fong2016, Bandurin2016,Levitov2016, Kumar2017, Bandurin2018,Berdyugin2019, Andreev2022N}  and experimental \cite{Molenkamp1994, Molenkamp1995, Moll2016,Gooth2018, Gusev2018,Bakarov2018,Gusev2021,Keser2021,Ella2019,Sulpizio2019,Kumar2022,Ahron2022} advances.\\

\begin{figure}[!t]
		\centering
		\includegraphics[width=\columnwidth]{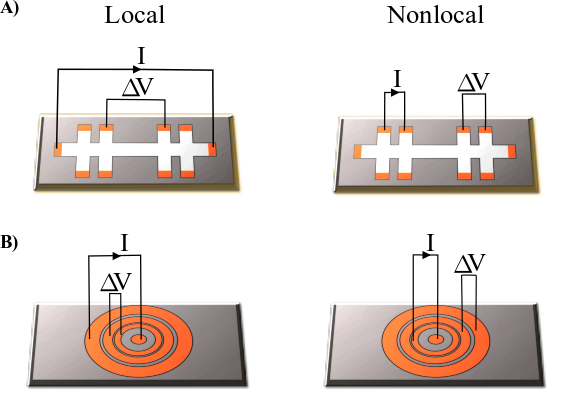}
		\caption {{Schematics of Hall bar and multi-terminal Corbino geometry.} 
			Measurement lead configurations for local and nonlocal measurements are shown in  ({\bf A}) for the Hall bar and ({\bf B}) the Corbino geometry.}
		\label{fig:Fig1}
	\end{figure}

	{\it Corbino samples parameters and experimental protocol}.---Typical measurements performed in large 2DEG samples utilize the four-terminal setup either in {van der Pauw (VdP)} or Hall bar geometries, and the latter is shown in Fig.\ref{fig:Fig1}A for both local and nonlocal measurement configurations. The Corbino geometry shown in Fig.\ref{fig:Fig1}B was chosen to ensure that only the bulk of the 2DEG was probed. To avoid any contributions from the Ohmic contacts, four-terminal Corbino devices were patterned on two symmetrically doped GaAs/AlGaAs heterostructures, one with a $40~nm$ (CBM302) and the other with a $30~nm$ (CBM301) quantum well widths, see Supplementary Material (SM). The devices have electron densities of  $n_e=1.7\times 10^{11}$ $cm^{-2}$ and  $n_e=3.6 \times 10^{11}$ $cm^{-2}$, and electron mobilities of  $20.3 \times 10^{6}$ $cm^2 (Vs)^{-1}$ and $27.8 \times 10^{6}$ $cm^2 (Vs)^{-1}$ below 20 $mK$, respectively. Both Corbino devices consist of three 2DEG active rings with inner/outer radii of  $(150/750)~\mu m, (960/1000)~\mu m$, and $(1300/1400)~\mu m$ respectively. These annular regions are defined by Ge/Au/Ni/Au layers of $26/54/14/100~nm$ thickness that diffuses into the wafer during an annealing process. As a result, Ohmic contacts  with resistances on the order of $\sim 3~\Omega$ are formed. \eendparagraph% 
	
	From their electron densities the $e-e$ interaction parameter $r_s=(\pi n_e a_B^2)^{-\frac{1}{2}}$, where $a_B$ is the effective Bohr radius, were calculated to be $1.32$ for CBM302 and $0.92$ for CBM301, and from their electron mobilities $\ell_{MR}$ values were calculated to be $138~\mu m$ for CBM302 and $271~\mu m$ for CBM301. The device with a lower electron density, CBM302, has a higher $r_s$ parameter, indicating an enhanced $e-e$ interaction compared to CBM301. The multi-terminal Corbino devices were cooled down to the base temperature $T \simeq 6~mK$ of a dilution refrigerator, illuminated by a red LED from room temperature to approximately $T \simeq 7~K$ to optimize the electron mobility. All nonlocal measurements presented in this work were performed with a fixed current of $100~nA$ applied to the two innermost rings, and the voltage difference was measured across the outermost rings.\\

\begin{figure}[!t]
		\centering
		\includegraphics[width=\columnwidth]{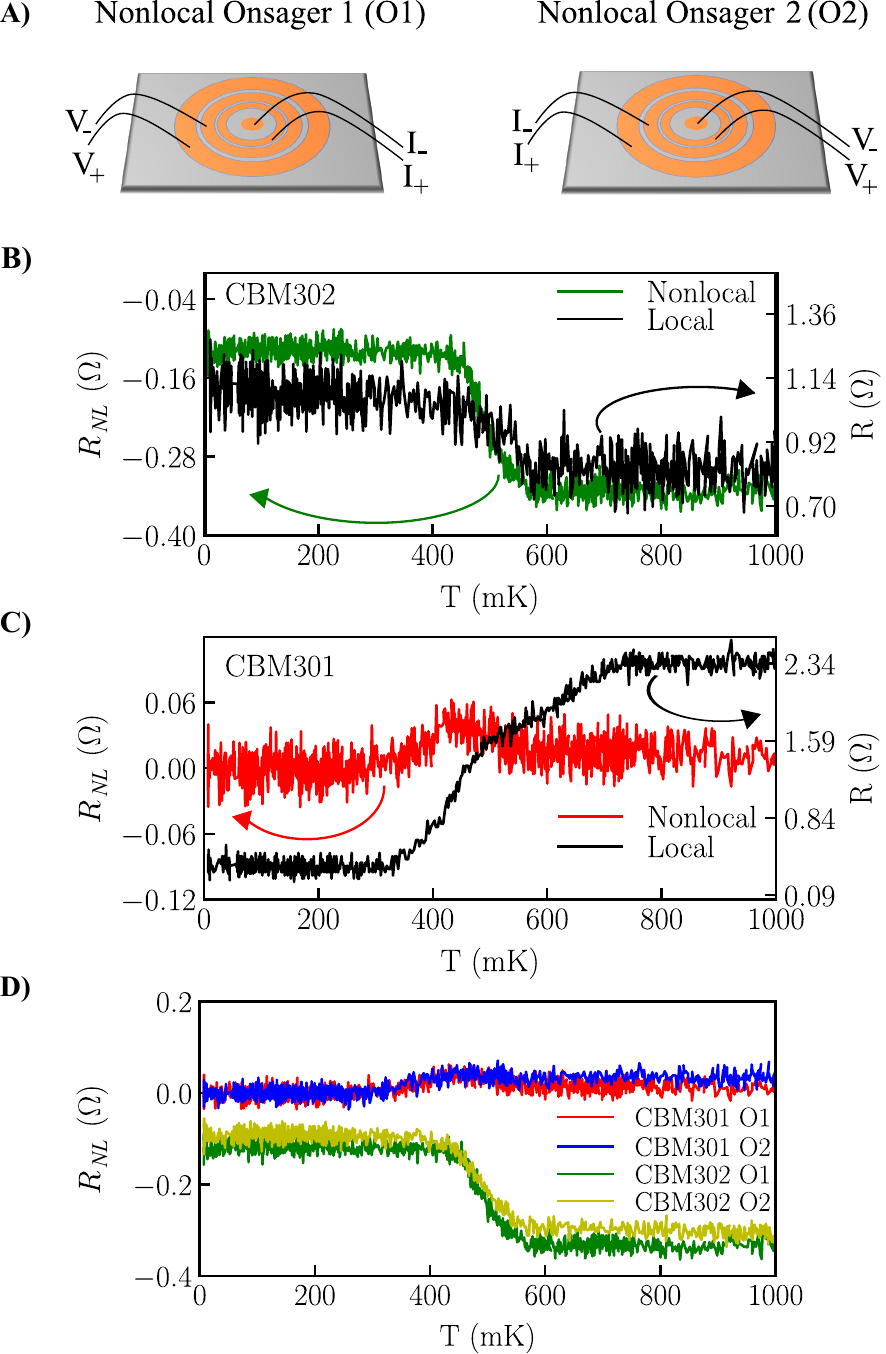}
		\caption{{Local and nonlocal transport measurements in multi-terminal Corbino 2DEGs.} 
			({\bf A}) Schematic representation of two distinct four-point nonlocal measurement configurations.
			({\bf B}) Nonlocal resistance measured in O1 configuration is compared with local measurements for CBM302 and
			({\bf C}) for CBM301. ({\bf D}) Nonlocal measurements in CBM301 and CBM302 for both O1 and O2 configurations. Onsager reciprocity  is validated within experimental resolution.
		}
		\label{fig:Fig2}
	\end{figure}

	{\it Local and nonlocal measurements}.---To ensure consistency, we performed both local and nonlocal measurements in both multi-terminal Corbino devices during the same cooldown, and the results are shown in Fig.\ref{fig:Fig2}B and C. For the nonlocal measurements, a current $I$ was applied to the two inner rings and the voltage difference $\Delta V_{NL}$ was measured across the two outer rings{ (the Onsager 1 (O1) configuration)}, and vice-versa for the Onsager 2 (O2) configuration, shown in Fig.\ref{fig:Fig2}A. The temperature dependence of nonlocal resistance $R_{NL}\equiv \Delta V_{NL}/I$ is shown in Fig.\ref{fig:Fig2}B by a green line for CBM302 and by a red line for CBM301 in  Fig.\ref{fig:Fig2}C, from base temperature ($6~mK$) to $1~K$. In CBM302, a nearly constant negative nonlocal signal was measured up to $T\simeq 450~mK$, at which point a sharp {increase} of the negative signal occurs concomitantly with the sharp decrease in local resistance. Between temperatures $T=450~mK$ and $T=1~K$, our measurements in CBM302 demonstrate a ${\sim}160\%$ increase in negative nonlocal signal, and a ${\sim}33\%$ decrease in local signal. As discussed below, these two observations are inconsistent with ballistic effusive electronic transport \cite{Leonid2018,Gupta2021}, and rather point towards the formation of a continuum and electronic flow governed by hydrodynamics when the Knudsen number $\zeta\equiv l_{MC}/l_{MR}\lesssim 1$, as is the case here. Consistent with our previous report \cite{Sujatha2023}, CBM301 shows an opposite behaviour to CBM302, and within experimental resolution the nonlocal resistance is absent except for a small positive hump occurring at $T\simeq 400$ $mK$. \\

	{\it Nonlocal reciprocity measurements}.---To test the reciprocity relations, we measured the nonlocal response for both O1 and O2 configuration, see Fig.\ref{fig:Fig2}A. Despite the smaller current-driving area and larger measurement area in the O2 configuration compared to O1, we observe no notable differences in nonlocal resistance and their temperature dependence is shown in Fig.\ref{fig:Fig2}D. Clearly, the Onsager reciprocity relations are satisfied over the temperature range probed in our experiment, {\it i.e.} from {$T=6~mK$} to $T=1~K$, as expected in hydrodynamic transport \cite{Olmstead1975}. This also shows that the nonlocal signal is not caused by an accumulation of charges, as reciprocity measurements should have been different if the electron density remains constant between the inner and outer annular rings that have  significantly different areas. In the SM, we show that the reciprocity relations are analytically satisfied for our system based on the hydrodynamic model presented in the following section. Finally, we performed probe symmetry measurements between adjacent contacts, also shown in the SM, which are all proved to be satisfied.\eendparagraph

	{\it Electron collision and kinetic theory}.---In a continuum, the role played by individual particle states diminishes and the flow becomes governed by the transport of a macroscopic fluid ensemble of particles. When this is the case, the details of their macroscopic behaviour arise from the description of individual particle states through solving Boltzmann's Transport Equation (BTE) of kinetic theory and determining macroscopic quantities. For electron flow however, interparticle interactions become important, requiring higher-order terms for an accurate description.
	Macroscopically, these interactions are described as body forces acting on the collective behaviour of the system.
	In our case, the interaction mechanisms arise by scattering, and we consider two main forms: momentum conserving $e-e$ scattering and momentum relaxing electron-impurity scattering, with relaxation times $\tau_{MC}$ and $\tau_{MR}$ respectively. We also omit the contribution to momentum relaxing due to electron-phonon scattering in our modelling since  it has been  shown to play a negligible role below a temperature of $1~K$ \cite{Sarma1992}. Moreover, it is known that well below $1~K$,  electron-impurity scattering does not depend on temperature, whereas electron-electron has an inverse square power relationship \cite{Sarma2022,Quinn1982}. Despite the common misconception that momentum-conserving interactions do not affect bulk resistivity, in a Corbino device without edges our observations of a non-monotonic temperature dependent transport and the realization of Gurzhi effect must relate to momentum conserving scattering.
	As such, when deriving the collective behaviour from BTE, we must consider the effects of a collision integral that is non-zero under velocity space integration.\eendparagraph
	
	In the presence of an electric field $\vec{E}$, the linearized transport equation is \cite{Ziman1960}
	\begin{equation}\label{eq:BoltzmannTransport}
		\frac{\partial f_{\vec{v}}}{\partial t} + \vec{v}\cdot\frac{\partial f_{\vec{v}}}{\partial \vec{r}} + \frac{q_e}{m^*} \vec{E}\cdot\frac{\partial f^0_{\vec{v}}}{\partial \vec{v}} = \left(\frac{\partial f_{\vec{v}}}{\partial t}\right)_{coll},
	\end{equation}
	where $(\vec{v},\vec{r},t)$ is particle velocity and position at time $t$, $f_{\vec{v}}(\vec{v},\vec{r},t)$ is the electron distribution function, $f^0_{\vec{v}}(\vec{v},\vec{r},t)$ is the equilibrium distribution in the absence of an electric field, $q_e$ is electron charge $-e$ and $m^*$ is effective mass ($0.067 m_e$ for GaAs/AlGaAs). Using the relaxation time approximation, momentum relaxing scattering relaxes the system to a field-less equilibrium, whereas momentum conserving scattering relaxes the system to an equilibrium where electrons drift at velocity $\vec{u}$ with distribution function $f^0_{\vec{u}}$. The collision integral then becomes
	\begin{equation}\label{eq:CollisionIntegral}
		\left(\frac{\partial f_{\vec{v}}}{\partial t}\right)_{coll} = -\frac{f_{\vec{v}} - f^0_{\vec{v}}}{\tau_{MR}} - \frac{f_{\vec{v}} - f^0_{\vec{u}}}{\tau_{MC}}.
	\end{equation}
	For small drift velocity, expanding $f^0_{\vec{u}} = f^0_{\vec{v}}\left(\epsilon - m^*\vec{v}\cdot\vec{u}\right)$ around static equilibrium yields
	\begin{equation}\label{eq:DriftEquilibrium}
		f^0_{\vec{u}} \approx f^0_{\vec{v}} - 2 \vec{u} \cdot \frac{\partial f^0_{\vec{v}}}{\partial \vec{v}},
	\end{equation}
	and substituting this result into Eq.\ref{eq:CollisionIntegral} gives
	\begin{equation}
		%\begin{eqnarray}
		\left(\frac{\partial f_{\vec{v}}}{\partial t}\right)_{coll} = %&=&
		-\frac{f_{\vec{v}} - f^0_{\vec{v}}}{\tau_T} - \frac{2}{\tau_{MC}}\vec{u}\cdot\frac{\partial f^0_{\vec{v}}}{\partial \vec{v}},
	\end{equation}
	%\end{eqnarray} 
	where $\tau^{-1}_T = \tau^{-1}_{MC} + \tau^{-1}_{MR}$ is the total mean-free time of all scattering events by Matthiessen's rule. This result is consistent with previous work in lower dimensions \cite{Molenkamp1994,Kumar2022,Moore2017} and finally Boltzmann's equation now reads
	\begin{equation}\label{eq:finalBoltzmann}
		\frac{\partial f_{\vec{v}}}{\partial t} + \vec{v}\cdot\frac{\partial f_{\vec{v}}}{\partial \vec{r}} + \left(\frac{q_e}{m^*} \vec{E} + \frac{2\vec{u}}{\tau_{MC}}\right)\cdot\frac{\partial f^0_{\vec{v}}}{\partial \vec{v}} = - \frac{f_{\vec{v}} - f^0_{\vec{v}}}{\tau_T}.
	\end{equation}
	By taking the first velocity moment of Eq.\ref{eq:finalBoltzmann}, we arrive at the Navier-Stokes equation where the third and fourth term give body force with the following scattering corrections:
	\begin{eqnarray}
		\vec{f} &=& \frac{q_e}{m^*}\vec{E} + \vec{u}\left(\frac{2}{\tau_{MC}} - \frac{1}{\tau_T}\right)\nonumber \\ 
		&=& -\frac{q_e}{m^*}\vec{\nabla}\phi + \frac{\vec{u}}{\tau},\label{eq:bodyForce}
	\end{eqnarray}
	where $\tau^{-1} = \tau_{MC}^{-1}-\tau_{MR}^{-1}$ and $\vec{E} = - \vec{\nabla} \phi$. This macroscopic description of the body force originating from the single-particle picture will be shown below to be of paramount importance for the nonlocal electronic transport observed  in the hydrodynamic regime.\eendparagraph
	
	{\it Hydrodynamics and electron viscosity}.---The Navier-Stokes equation of  hydrodynamics for incompressible flow can be written as
	
	\begin{eqnarray} \label{eq:ns_continuity}
		\nabla \cdot \vec {u}  &=& 0, \\
		\frac{\partial\vec{u}}{\partial t} +\left(\vec{u} \cdot \vec{\nabla}\right) \vec{u} &=& \nu \vec{\nabla}^2\vec{u} + \vec{f}.\label{eq:ns_momentum}
	\end{eqnarray}
	In our case, $\vec{u}$ describes the electron drift velocity, $\nu = \frac{v_f \ell_{MC}}{4}$ is the electron viscosity \cite{Polini2015}  with $v_f$ the Fermi velocity, and $\vec{f}$ is the body force per unit mass
	given by Eq.\refeq{eq:bodyForce}. For a steady state flow ($\frac{\partial\vec{u}}{\partial t}=0$), we can re-write Eq.\refeq{eq:ns_momentum} as
	\begin{equation}\label{eq:ns_steadystate}
		\left(\vec{u} \cdot \vec{\nabla}\right) \vec{u} = \nu \vec{\nabla}^2\vec{u} -\frac{q_e}{m^{*}} \vec{\nabla} \phi + \frac{\vec{u}}{\tau}.
	\end{equation}
	
	At zero magnetic field, $\vec{u} = (u_r,0,0)$ in polar coordinates for the Corbino geometry \cite{corbinotron}. As shown in the SM, analytic solutions to Eq.\ref{eq:ns_continuity} and Eq.\ref{eq:ns_steadystate} give the general forms of $u_r$ and $\phi$ as
	
	\begin{equation}\label{eq:u_analytic}
		u_r \propto \frac{1}{r},\hspace{12pt} \phi \propto \tau^{-1}\log(r), 
	\end{equation}
	and as a result when the steady-state solution of $u_r$ is substituted into Eq.\refeq{eq:ns_momentum}, the viscosity term $\nu \vec{\nabla}^2 \vec{u}$ disappears.
	Although this may seem unusual as it would suggest that viscosity plays no role at first glance, viscosity is responsible for resisting radial shear deformations, where fluid layers move at different speeds along the direction of motion. This shear deformation occurs due to radial velocity gradients, such as those seen in the discontinuous motion at contact boundaries in Ohmic transport, and causes flow to extend beyond the discontinuity. It is only at this point that the viscosity term truly does drop out. As flow beyond the boundary is  is a direct result of this viscosity term, one would expect an explicit dependence of $\nu$ in the electric potential profile $\phi$. In our model, this manifests as the dependence of $\ell_{MC}$ in $\tau$, that is directly related to $\nu$ and depends greatly on temperature below $1~K$. A distinct feature of radial hydrodynamic transport, viscous electron flow  can now extend beyond the source-drain region and be experimentally probed.\eendparagraph
	
		\begin{figure}[!ht]
		\centering
		\includegraphics[width=\columnwidth]{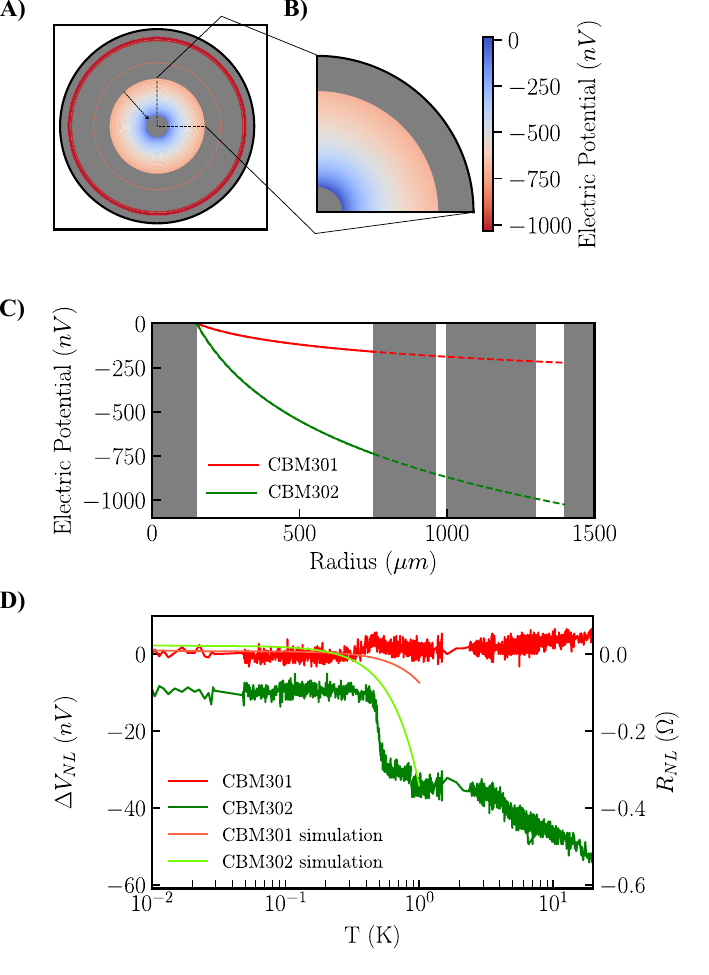}
		\caption{{Nonlocal simulation results.} Simulation results are shown  with the center contact as the grounding reference point.
			({\bf A}) The results of nonlocal simulation for CBM302 with shaded regions representing Ohmic contacts. ({\bf B}) A zoom in of the active simulation region.
			({\bf C}) Results for both CBM301 and CBM302 along the central axis, with the dotted lines indicated the region of extrapolation. Ohmic contacts are shaded grey.
			({\bf D}) Comparison with simulated and experimental results for both devices is shown.}
		\label{fig:Fig3}
	\end{figure}
	
	{\it Navier-Stokes simulation and comparison with experiment}.---To confirm our findings, we modelled the electronic system using the Navier-Stokes equations \ref{eq:ns_continuity} and \refeq{eq:ns_momentum}. We performed simulations of these equations for nonlocal configurations in CBM301 and CBM302 from base temperature to $T=1~K$. In this configuration, simulations are extended from the active simulation region (the current path) to the outer rings through extrapolation, by fitting drift velocity $\vec{u}$ and electric potential $\phi$ to general functions resembling Eq.\refeq{eq:u_analytic}. This extrapolation technique is verified by simulating the entire region, and is shown in section 5.3 of SM.
	To obtain the voltage difference, we subtracted electric potential from edge-to-edge of each contact, using the configuration shown in Fig.\ref{fig:Fig2}A O1. As the ohmic contacts are made by diffusion, the metal-rich phases penetrate the heterostructure and one may wonder if it obstructs the 2DEG. In an extensive study of diffusion by way of  TEM microscopy and EDX spectroscopy, it was concluded \cite{Koop2013} that the annealing produces a highly doped region within the 2DEG making the Schottky barrier very thin. As a result, 2DEGs  with heterostructures similar to  those presented  in our work can retain their electronic transport characteristics. While this obstruction likely induces disorder to some unknown degree, and could modify the profile of $\phi$ within the contact region, by taking the edge-to-edge potential difference, the overall effect of the obstruction drops out. \\
	
	The temperature dependence of simulated values of the nonlocal potential are shown in Fig.\ref{fig:Fig3}D together with the experimental values of potential difference measured. While the simulation does not capture the negative nonlocal potential observed below $\sim 300$ mK temperature in the super ballistic transport regime, it nevertheless semi-quantitatively explains the strong (weak) nonlocal voltages observed in CBM302 (CBM301) between 450 mK and 1~K. Furthermore, the obtained simulation value of  $-7.4\pm{0.1}~nV$ for CBM301 at $T=1~K$ confirms the near-zero nonlocal resistance in our experimental data is not a result of dominant hydrodynamic transport.  Finally, for temperatures well above $1~K$, we verified that including phonon contributions to resistances  by adding experimentally determined electron-phonon scattering rates to our model \cite{Mittal1996} restored the expected behavior, {\it i.e.} a monotonic increase in resistance towards a positive value.\eendparagraph

	{\it Conclusion}.---We have observed  non-zero local and nonlocal voltages found to be akin to a Gurzhi effect, albeit in the bulk in a 2DEG Corbino device with high $e-e$ interaction that meets the length scale criteria for hydrodynamic transport, {\it i.e. at Knudsen number $\zeta \lesssim 1$}.  The nonlocal signal was found to be nearly absent in a lower $e-e$ interaction device that failed to meet the same length scale criteria in a similar temperature range of investigations. Due to the bulk nature of electronic transport in the Corbino geometry and considering the Navier-Stokes equations, we conclude that the body force of carriers plays an integral role in dictating the flow of electrons, and necessarily contains the electron viscosity. This is turn allows viscous drag flow to extend beyond the contact boundary, as is observed in the experiment. Looking forward, our confirmation of hydrodynamic transport in GaAs/AlGaAs Corbino rings may allow the probing of Hall viscosity, an intriguing dissipationless bulk property arising from  the odd symmetry of the viscosity tensor in two dimensions \cite{Tobias2019}.\eendparagraph

	{\it Acknowledgments}.---This work has been supported by NSERC (Canada), FRQNT-funded strategic clusters INTRIQ (Québec) and Montreal-based CXC. The Princeton University portion of this research is funded in part by the Gordon and Betty Moore Foundation’s EPiQS Initiative, Grant GBMF9615.01 to Loren Pfeiffer. Sample fabrication was carried out at the McGill Nanotools Microfabrication facility. We would like to thank T. Szkopek and F. Boivin for the insightful comments on the manuscript, B.A. Schmidt and K. Bennaceur for their technical expertise during the fabrication and earlier characterization of the Corbino sample, and R. Talbot, R. Gagnon, and J. Smeros for general laboratory technical assistance. This work was performed, in part, at the Center for Integrated Nanotechnologies, an Office of Science User Facility operated for the U.S. Department of Energy (DOE) Office of Science. Sandia National Laboratories is a multimission laboratory managed and operated by National Technology \& Engineering Solutions of Sandia, LLC, a wholly owned subsidiary of Honeywell International, Inc., for the U.S. DOE’s National Nuclear Security Administration under contract DE-NA-0003525. The views expressed in the article do not necessarily represent the views of the U.S. DOE or the United States Government.\eendparagraph
	
	{\it Authors contributions}. S.V. and G.G. conceived the experiment.  K.W.W. and L.N.P. performed the semiconductor growth by molecular beam epitaxy and provided the material. S.V. fabricated  the Corbino samples, conducted the electronic transport measurements at low temperatures and performed the data analysis. Z.B.K. performed the theoretical modelling and the numerical simulations, assisted by J. M. and S.V. Important expertise and insights on semiconductor physics and data interpretation were provided by L.W.E and M.P.L. The manuscript was written by S.V., Z.B.K., and G.G, and all authors commented on it. Finally, both S.V. and Z.K.B contributed equally to this work.\\
	
	%\end{document}
	\nocite{*}
	%\bibliography{mybib}
	%\\
	%\\
	\bibliographystyle{apsrev4-1}

\pagebreak

\widetext
\newpage
\begin{center}
	\textbf{\large Supplementary Material}
\end{center}

%%%%%%%%%% Merge with supplemental materials %%%%%%%%%%
%%%%%%%%%% Prefix a "S" to all equations, figures, tables and reset the counter %%%%%%%%%%
\setcounter{equation}{0}
\setcounter{figure}{0}
\setcounter{table}{0}
\setcounter{page}{1}
\makeatletter
\renewcommand{\theequation}{S\arabic{equation}}
\renewcommand{\thefigure}{S\arabic{figure}}
\renewcommand{\bibnumfmt}[1]{[S#1]}
\renewcommand{\citenumfont}[1]{S#1}
%%%%%%%%%% Prefix a "S" to all equations, figures, tables and reset the counter %%%%%%%%%%
% Define equation referencing
%\usepackage[hscale=0.7,vscale=0.8]{geometry}
\def\equationautorefname~#1\null{(#1)\null}
\newcommand{\smod}{\hspace{-0.3cm}\mod}
%\doublespacing
\renewcommand\thesection{\arabic{section}}
\renewcommand\thesubsection{\thesection.\arabic{subsection}}
\newpage

\section {Heterostructure}
\begin{figure}[!ht]
	\centering
	\includegraphics[width=1.1\textwidth]{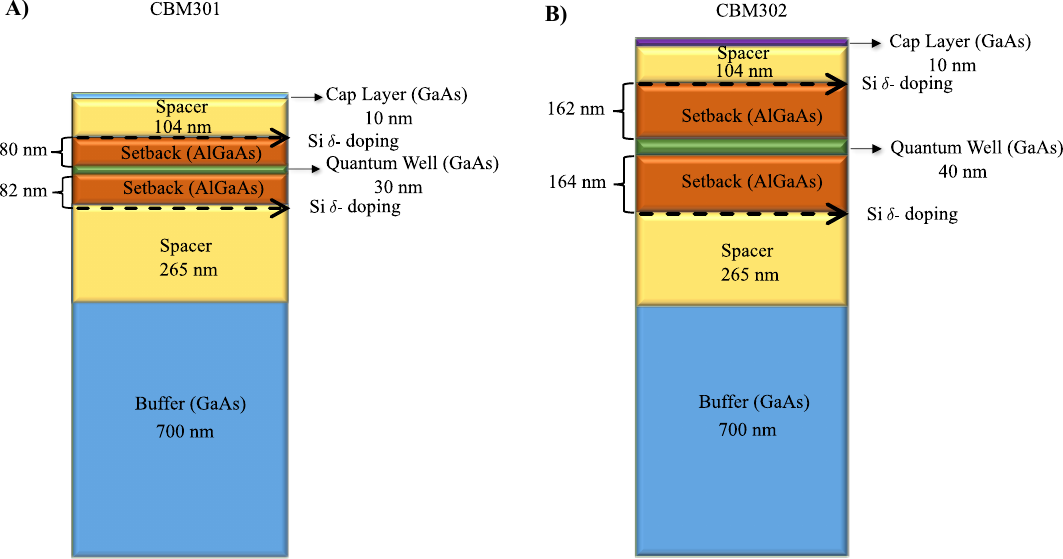}
	\caption{Heterostructure of wafers for sample (A.) CBM301  and (B.) CBM302. The figure indicates the materials used and the layer thickness.}
	\label{fig:FigS1}
\end{figure}

The devices used in the manuscript were fabricated on two different wafers, each with a different heterostructure as shown in Fig.\ref{fig:FigS1}. These two wafers have GaAs substrates that are $700~nm$ thick. The heterostructure is deposited on top of the substrate, with spacer material acting as a buffer between layers. The distance between the dopant and the edge of the quantum well is denoted as the setback distance. Finally, the heterostructure is protected by a cap layer made of a material similar to the substrate.\\

\section {Measurement circuits}

\subsection{Local measurement Circuit}
\begin{figure}[!htb]
	\centering
	\includegraphics[width=0.85\textwidth]{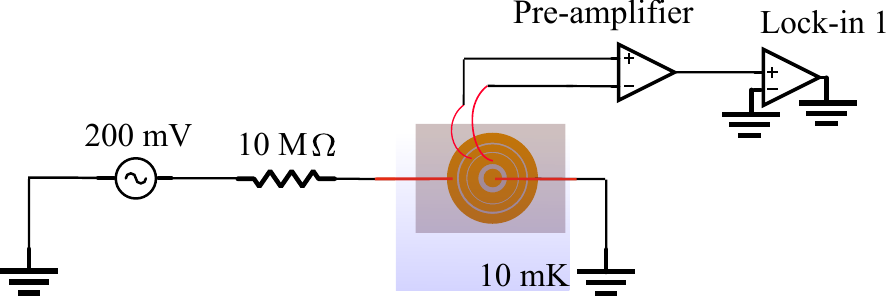}
	\caption{Local transport measurement circuit used to determine the four point resistance of the Corbino sample with a fixed current of 20 $nA$.}
	\label{fig:FigS2}
\end{figure}

The circuit in local transport measurements, depicted in Fig.\ref{fig:FigS2}, consists of an SR830 lock-in amplifier, a $10~M\Omega$ resistor and a SR560 pre-amplifier. It is a basic resistance measurement setup, where a fixed current is applied and the voltage difference is measured across the device. In this setup, an AC voltage of $200~mV$ from the SR830 is applied across the $10~M\Omega$ resistor connected in series with the Corbino sample. This configuration allows for a constant current of $20~nA$ to be applied to the outermost contact of the multi-terminal Corbino, while the innermost contact is kept grounded. The voltage difference across the intermediate contacts is measured and amplified using the SR560 pre-amplifier with a gain of 100.\\

\vspace*{-12pt}
\subsection {Nonlocal measurement circuit}

\begin{figure}[!htb]
	\centering
	\includegraphics[width=0.85\textwidth]{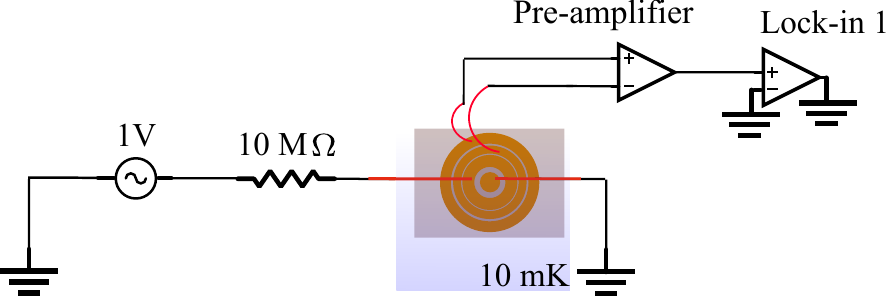}
	\caption{Nonlocal transport measurement circuit with a fixed current of $100~nA$.}
	\label{fig:FigS3}
\end{figure}

The nonlocal measurement circuit is  identical to the local measurement circuit described above and uses the same elements. Here, an AC signal of $1~V$ is transmitted through the $10~M\Omega$ resistor resulting in a $100~nA$ current as shown in Fig.\ref{fig:FigS3}. This electric current flows across the inner two contacts by keeping the innermost contact grounded, and vice versa for Onsager measurements. The voltage difference is measured across the two outer rings. To verify if any superconducting and/or proximity effects of the contacts contributed to our transport measurements, we conducted a DC-biased measurement with a $300~nA$ bias current. The obtained I-Vs were linear, {demonstrating Ohmic electronic transport}. \\

\subsection{Lock-in frequency}
{%\color{red}
	All transport measurements presented in the main text were conducted with lock-in signal frequencies of $73.51~\text{Hz}$ for CBM302 and $67.89~\text{Hz}$ for CBM301. To verify there was no frequency-dependent effects, we also performed the same measurements with lower lock-in signal frequencies of $13.957~\text{Hz}$ for CBM302 and $17.253~\text{Hz}$ for CBM301. The signal and its temperature dependence remains the same, albeit with smaller signal amplitudes, that we attribute to different cooldown conditions.\\

	\section {Probe symmetry}
	
	\begin{figure}[!htb]
		\centering
		\includegraphics[width=1.0\textwidth]{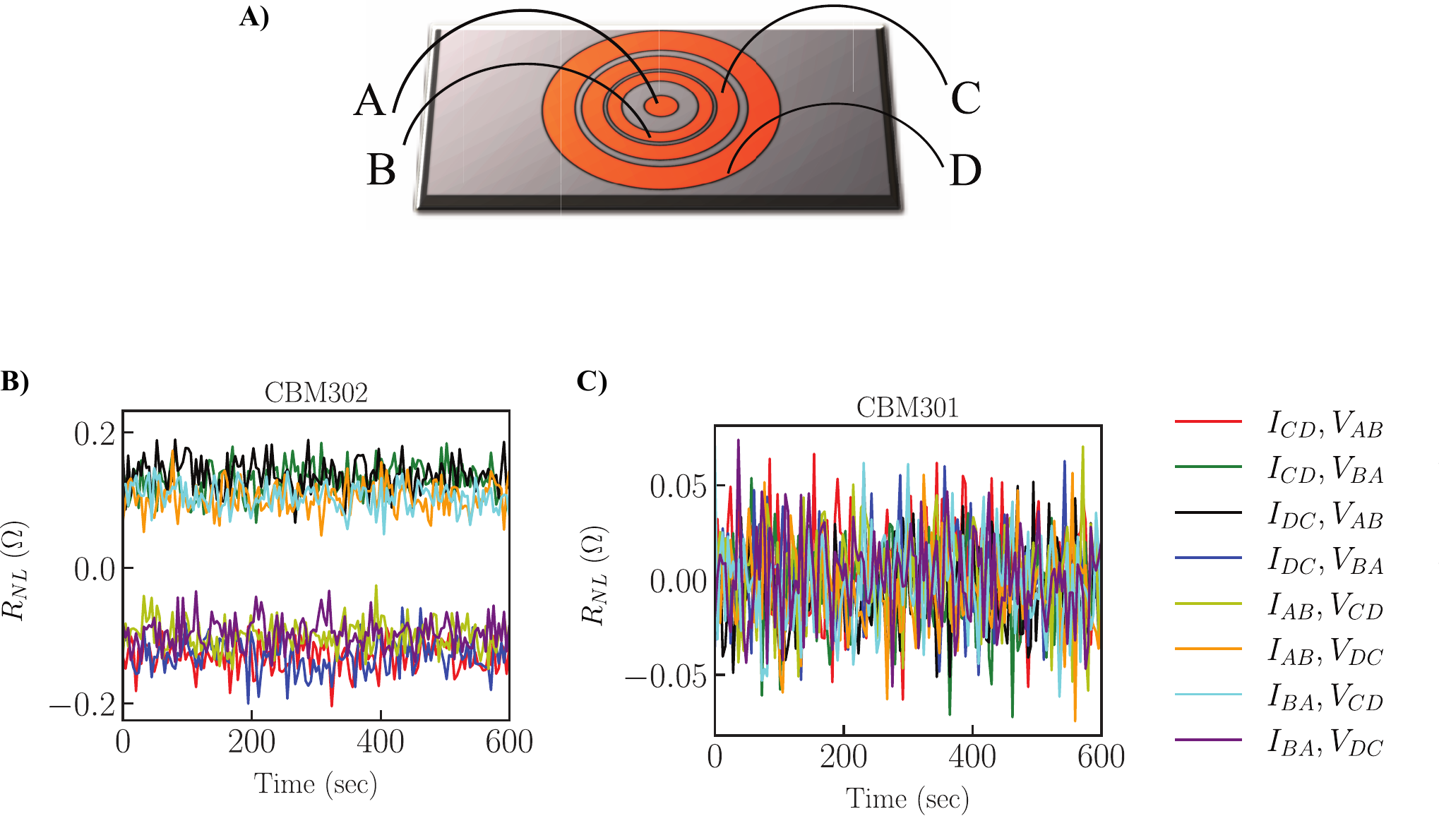}
		\caption{Panel ({\bf A}) A cartoon of the Corbino device with each contact labeled as A, B, C and D. ({\bf B}) and ({\bf C}) show the nonlocal resistance measurements under various probe configurations in CBM302 and CBM301 respectively. The legend indicates contact configurations, where $I$ is the applied current and V is the measured voltage difference.}
		\label{fig:probe_symmetry}
	\end{figure}
	
	To further confirm the results presented in Fig.2 of the main text, we also conducted nonlocal transport measurements using different probe configurations. The data obtained for both CBM301 and CBM302 at base temperature and zero magnetic fields ($B=0~T$) are shown in Fig.\ref{fig:probe_symmetry}. The results show that all probe symmetries are respected in both devices.

	\section{Navier-Stokes in hydrodynamics}
	
	\subsection{Length scales}
	As mentioned in the main text, the length scale conditions for hydrodynamic transport are $\ell_{MC}\ll W \ll \ell_{MR}$, where $W$ is channel width, $\ell_{MC}$ is momentum conserving scattering mean-free path (in our case the $e-e$ scattering), and $l_{MR}$ is momentum relaxing scattering mean-free path. At low temperatures ($T<1~K$) where phonon contribution is negligible, we consider only momentum relaxing scattering due to electron-impurity,
	\begin{equation}
		\ell_{MR} = 5.22 \mu \sqrt{n_e},
	\end{equation}
	where $\mu$ is carrier mobility and $n_e$ is electron density. In our previously reported study \cite{Sujatha2023}, we verified experimentally that the electron density remains constant below $1~K$, and therefore we consider $\ell_{MR}$ to be constant for both devices with scattering time $\tau_{MR} = \frac{v_f}{\ell_{MR}}$, where $v_f$ is the Fermi velocity.

	\noindent{The quantum} lifetime for electron-electron interaction was calculated using a well known theoretical expression from Guillani and Quinn \cite{Quinn1982}:
	\begin{equation}
		\frac{1}{\tau_{MC}} \approx \frac{\left(k_B T\right)^2}{2 \pi \hbar E_f}\left[ \log\left(\frac{E_f}{k_B T}\right) + \log\left(\frac{2 q_{tf}}{k_f}\right) + 1  \right],
	\end{equation}
	where $E_f$ $k_f$, $q_{tf}$ and $k_B$ are the Fermi energy, Fermi wave vector, Thomas-Fermi wave vector and Boltzmann constant, respectively. In our simplified model, we consider a total scattering length making use of Matthiessen's rule, which is the mean-free path of any scattering event, given by
	\begin{equation}
		\ell^{-1}_T = \ell_{MC}^{-1} + \ell_{MR}^{-1}.
	\end{equation}
	Hence the total lifetime $\tau_T$ can be written as
	\begin{equation}
		\tau^{-1}_T = \frac{v_f}{\ell_T} = v_f \left(\ell_{MC}^{-1} + \ell_{MR}^{-1}\right) = \left(\frac{v_f}{\ell_{MC}} + \frac{v_f}{\ell_{MR}}\right) = \tau_{MC}^{-1} + \tau_{MR}^{-1}.
	\end{equation}

	\subsection{Governing equations}
	
	The Navier-Stokes equations for an incompressible viscous fluid with negligible pressure gradient are given by \cite{landaufluid}:
	\begin{eqnarray}\label{eq:navierstokes1}
		\nabla \cdot \vec {u}  &=& 0, \\
		\label{eq:navierstokes2}
		\frac{\partial\vec{u}}{\partial t} +\left(\vec{u} \cdot \vec{\nabla}\right) \vec{u} &=& \nu \vec{\nabla}^2\vec{u} + \vec{f}.
	\end{eqnarray} 

\noindent{Where} $\vec{u}$ is electron velocity, $\nu$ is kinematic viscosity of the system given by $\nu = \frac{\hbar}{4m^*} \sqrt{2 \pi n_e}~\ell_{MC}$, and $\vec{f}$ is body force per unit mass. There are two main body forces that we consider: the electromagnetic (external) and momentum changes due to scattering (internal). As shown in the main text, the total body force per unit mass if given by

	\begin{equation}\label{eq:6}
		\vec{f} = \vec{f}_{EM} + \vec{f}_{scat} = -\frac{q_e}{m^{*}} \vec{\nabla} \phi + \frac{\vec{u}}{\tau},
	\end{equation}
	where $q_e$ is electron charge $-e$, $m^*$ is effective mass ($0.067 m_e$ for GaAs/AlGaAs), $\tau$ is the effective mean-free scattering time $\tau^{-1}=\tau_{MC}^{-1}-\tau_{MR}^{-1}$, and $\phi$ represent the electric potential created by electron flow directed by an applied current.

	\subsection{Two-dimensional geometries}
	In a multiring Corbino device, the 2DEG is formed by an annulus, hence the natural description of this system is in polar coordinates $(r, \theta)$, with the center contact as the origin. In the absence of a magnetic field, the symmetries of our system do not permit flow in the $\theta$ direction, as our current carries electrons along the $r$-axis \cite{corbinotron}. As such, $\vec{u} = \left(u_r, 0, 0 \right)$ and $\phi(r,\theta) = \phi(r)$, and as a result equations \ref{eq:navierstokes1} and \ref{eq:navierstokes2} simplify to:
	\begin{eqnarray}
		\label{eq:navier2d1}
		\frac{1}{r} \frac{\partial\left(r u_r\right)}{\partial r} &=& 0, \\
		\label{eq:navier2d2}
		\frac{\partial u_r}{\partial t} + u_r\frac{\partial u_r}{\partial r} &=& \nu \left(\frac{1}{r}\frac{\partial}{\partial r}\left(r \frac{\partial u_r}{\partial r}\right)-\frac{u_r}{r^2}\right)-\frac{q_e}{m^*}\frac{\partial \phi}{\partial r}+\frac{u_r}{\tau},
	\end{eqnarray}
	which in the steady state ($\frac{\partial u_r}{\partial t}\rightarrow0$) has solutions:
	\begin{eqnarray}
		\label{eq:navier2dsoln1}
		u_r(r) &=& \frac{K}{r}\\
		\label{eq:navier2dsoln2}
		\phi(r) &=& \frac{K^2}{2}\frac{m^*}{q_e}\left(\frac{1}{r^2}-\frac{1}{r_{gnd}^2}\right) + \frac{K}{\tau}\frac{m^*}{q_e}\log\left(\frac{r}{r_{gnd}}\right)\nonumber\\
		&=& \frac{K}{\tau}\frac{m^*}{q_e}\log\left(\frac{r}{r_{gnd}}\right) + \mathcal{O}(K^2),
	\end{eqnarray}
	where $r_{gnd}$ is the reference position of the grounded contact edge. Noting that 2D current density is given by
	\begin{equation}\label{eq:10}
		J_r(r) = \frac{I}{2 \pi r} = n_e q_e {u}_r = n_e q_e \frac{K}{r},
	\end{equation}
	we find that
	\begin{equation}\label{eq:constantK}
		\left|K\right| = \left|\frac{I}{2 \pi n_e q_e}\right| \ll 1,
	\end{equation}
	where the sign of $I$ indicates the direction of current.
	
	\subsection{Hydrodynamic reciprocity} 
	In the realm of viscous hydrodynamic transport it is known that the reciprocity relation is satisfied \cite{Olmstead1975} and the reciprocity identity for two different particle states, with a velocity $\vec{u}^{(1)}$ at position $r^{(1)}$ and $\vec{u}^{(2)}$ at position $r^{(2)}$ is
	\begin{equation}\label{eq:recipidentity}
		\vec{f}^{(1)} \cdot { \vec{u}^{(2)} }= \vec{f}^{(2)} \cdot \vec{u}^{(1)},
	\end{equation}
	
	\noindent{where} $\vec{f}^{(1)}$ and $\vec{f}^{(2)}$ are the corresponding body forces. In our case these two different states, labeled ${(1)}$ and ${(2)}$, denote the two Onsager configurations O1 and O2. The experimental evidence of satisfied reciprocity relations is presented in Fig.2D of the main text. Using the solutions in polar coordinates given by Eq.\ref{eq:navier2dsoln1}-\ref{eq:navier2dsoln2} and the body force given by  Eq.\ref{eq:6}, the reciprocity identity becomes
	\begin{eqnarray}
		\label{eq:recip_LHS}
		\vec{f}^{(1)}\cdot\vec{u}^{(2)} &=& -\left(\frac{K}{\tau}\frac{m^*}{q_e}\frac{\partial}{\partial r}\log\left(\frac{r^{(1)}}{r_{gnd}}\right) - \frac{1}{\tau}\frac{K}{r^{(1)}}\right) \left(\frac{K}{r^{(2)}}\right)\nonumber\\
		&=& \frac{1}{r^{(1)}r^{(2)}}\frac{K^2}{\tau}\left(1-\frac{m^*}{q_e}\right),\\
		\label{eq:recip_RHS}
		\vec{f}^{(2)}\cdot\vec{u}^{(1)} &=& -\left(\frac{K}{\tau}\frac{m^*}{q_e}\frac{\partial}{\partial r}\log\left(\frac{r^{(2)}}{r_{gnd}}\right) - \frac{1}{\tau}\frac{K}{r^{(2)}}\right) \left(\frac{K}{r^{(1)}}\right)\nonumber\\
		&=& \frac{1}{r^{(2)}r^{(1)}}\frac{K^2}{\tau}\left(1-\frac{m^*}{q_e}\right).
	\end{eqnarray}
	
	\noindent{Thus }the hydrodynamic reciprocity relation $\vec{f}^{(1)} \cdot \vec{u}^{(2)} = \vec{f}^{(2)} \cdot \vec{u}^{(1)} $ is satisfied. This support our experimental observation of nonlocal resistance using distinctive Onsager probe configurations.
	
	\section{Simulation}
	Numerical simulations of Eq.\ref{eq:navier2d1}-\ref{eq:navier2d2} were performed in Python via Finite-Difference Method (FDM) in the 2D polar coordinate system. In FDM, first-order and second-order spatial derivatives are discretized using the central difference approximation,
	
	\vspace*{-12pt} 
	\begin{align}\label{eq:centraldiff_firstorder_general}
		\frac{\partial f}{\partial x} &\approx \frac{f(x+h) - f(x-h)}{2h} + \mathcal{O}(h^2),\\
		\label{eq:centraldiff_secondorder_general}
		\frac{\partial^2 f}{\partial x^2} &\approx \frac{f(x+h) + f(x-h) - 2 f(x)}{h^2} + \mathcal{O}(h^2).
	\end{align}
	We employ a 2D uniformly spaced mesh grid for axes ($r$,$\theta$) with dimension $(N_r, N_\theta)$ and constant spacing $\Delta r$ and $\Delta \theta$. As the $\theta$ axis is periodic, the spatial derivatives take the following form:
	
	\vspace*{-12pt} 
	\begin{align}
		\label{eq:dfdr_sim}
		\frac{\partial f_{i,j}}{\partial r} &= \frac{f_{i+1,j} - f_{i-1, j}}{2 \Delta r},\\
		\label{eq:dfdtheta_sim}
		\frac{\partial f_{i,j}}{\partial \theta} &= \frac{f_{i,(j+1)\smod N_\theta} - f_{i, (j-1)\smod N_\theta}}{2 \Delta \theta},\\
		\label{eq:d2fdr2_sim}
		\frac{\partial^2 f_{i,j}}{\partial r^2} &= \frac{f_{i+1,j} + f_{i-1, j} - 2 f_{i,j}}{(\Delta r)^2},\\
		\label{eq:d2fdtheta2_sim}
		\frac{\partial^2 f_{i,j}}{\partial \theta^2} &= \frac{f_{i,(j+1)\smod N_\theta} + f_{i, (j-1)\smod N_\theta} - 2 f_{i,j}}{(\Delta \theta)^2},
	\end{align}
	where mod denotes the modulus operator.
	
	\subsection{Iterative solution}
	
	Simulations were performed iteratively over time, with time step $\Delta t$, using the approximation
	\begin{equation}\label{eq:dudt_forwarddifference}
		\frac{\partial u}{\partial t} \approx \frac{u^{t+1} - u^{t}}{\Delta t} +\mathcal{O}(\Delta t).
	\end{equation}
	Substituting into Eq.\ref{eq:navierstokes2}, we obtain
	
	\begin{equation}
		\frac{\vec{u}^{t+1} - \vec{u}^{t}}{\Delta t} = \nu \vec{\nabla}^2\vec{u}^t - \left(\vec{u}^t \cdot \vec{\nabla}\right) \vec{u}^t + \vec{f}^t,
	\end{equation}
	and solving for $\vec{u}^{t+1}$, we derive the update rule for velocity field at each iteration:
	\begin{equation}
		\vec{u^{t+1}} = \vec{u}^t + \Delta t \left( \nu \vec{\nabla}^2\vec{u}^t - \left(\vec{u}^t \cdot \vec{\nabla}\right) \vec{u}^t + \vec{f}^t \right).
	\end{equation}
	Factoring in for the body forces and substituting Eq.\ref{eq:6} yields
	
	\begin{equation}\label{eq:u_next_time_step}
		\vec{u^{t+1}} = \vec{u}^t + \Delta t \left( \nu \vec{\nabla}^2\vec{u}^t - \left(\vec{u}^t \cdot \vec{\nabla}\right) \vec{u}^t -\frac{q_e}{m^{*}} \vec{\nabla} \phi^t + \frac{\vec{u}^t}{\tau} \right).
	\end{equation}
	To obtain a solution for $\phi^t$ concurrently, we define a ``guess'' solution at each iteration:
	\begin{equation}\label{eq:guess_u}
		\vec{u}^P \equiv \vec{u}^t + \Delta t \left( \nu \vec{\nabla}^2\vec{u}^t - \left(\vec{u}^t \cdot \vec{\nabla}\right) \vec{u}^t + \frac{\vec{u}^t}{\tau} \right).
	\end{equation}
	Substituting this into Eq.\ref{eq:u_next_time_step}, we obtain:
	\begin{equation}\label{eq:simulation_u_tplus1_with_guess}
		\vec{u}^{t+1} = \vec{u}^P - \Delta t \frac{q_e}{m^{*}} \vec{\nabla} \phi^t.
	\end{equation}
	A solution for $\phi^t$ is obtained by taking the divergence of both sides of Eq.\ref{eq:simulation_u_tplus1_with_guess}. Continuity equation (Eq.\ref{eq:navierstokes1}) imposes the condition that $\vec{\nabla}\cdot\vec{u}^{t+1} = 0$, however this condition may not necessarily be true for our guess velocity $\vec{u}^P$ (Eq.\ref{eq:guess_u}). We are left with the following Poisson equation:
	
	\begin{equation}\label{eq:poisson_eqn_sim}
		\vec{\nabla}^2 \phi^t = \frac{1}{\Delta t }\frac{m^*}{q_e} \left(\vec{\nabla}\cdot\vec{u}^P\right).
	\end{equation}
	In 2D cylindrical coordinates, the Laplacian term of Eq.\ref{eq:poisson_eqn_sim} can be written as
	\begin{equation}\label{eq:poisson_term_2D}
		\vec{\nabla}^2 \phi^t = \frac{\partial^2 \phi^t}{\partial r^2} + \frac{1}{r}\frac{\partial \phi^t}{\partial r} + \frac{1}{r^2}\frac{\partial^2 \phi^t}{\partial \theta^2}.
	\end{equation}
	Using the discretized spatial derivative forms Eq.\ref{eq:dfdr_sim}-\ref{eq:d2fdr2_sim} of Eq.\ref{eq:poisson_term_2D} and solving for $\phi^t$ yields:

	\begin{multline}\label{eq:phi_discrete_full}
		\phi^t_{i,j} = \frac{1}{2} \left( \frac{\left(\Delta r\right)^2 \left(r \Delta \theta\right)^2}{\left(\Delta r\right)^2+\left(r \Delta \theta\right)^2} \right) \Biggl( \frac{\phi^t_{i+1,j}+\phi^t_{i-1,j}}{\left(\Delta r\right)^2} + 
		\frac{\phi^t_{i,(j+1)\smod N_\theta}+\phi^t_{i,(j-1)\smod N_\theta}}{\left(r \Delta \theta\right)^2} \\ +
		\frac{1}{r}\frac{\phi^t_{i+1,j}-\phi^t_{i-1,j}}{2\Delta r} -
		\frac{1}{\Delta t }\frac{q_e}{m^*} \left(\vec{\nabla}\cdot\vec{u}^P_{i,j}\right) \Biggr).
	\end{multline}
	Denoted as Poisson iterations, Eq.\ref{eq:phi_discrete_full} is calculated multiple times per time step. In our simulations, 100 Poisson iterations for each 5000 time steps were performed. The simulation converges to the steady-state solution of Eq.\ref{eq:navier2d1}-\ref{eq:navier2d2} when $\frac{\partial u^t}{\partial t} \rightarrow 0$. The convergence condition equivalently implies $\phi^{t}$ tends towards a constant value, which is shown in Fig.\ref{fig:FigS5} for CBM302 at $T=1~K$. Although we utilized the symmetries of our system to simplify equations Eq.\ref{eq:navierstokes1}-\ref{eq:navierstokes2} and derive analytic solutions Eq.\ref{eq:navier2dsoln1}-\ref{eq:navier2dsoln2}, our simulation framework avoids such approximations to retain general applicability.\\
	\newpage

	\subsection{Nonlocal simulations}

	For nonlocal simulations, we simulated only the region of the Corbino device with driven current, using the same experimental parameters. We then extrapolated our electric potential profile to the nonlocal region by curve fitting to a general function resembling analytic solution \ref{eq:navier2dsoln2}. Since we only simulate the active region of the 2DEG fluid (without Ohmic contacts), we measure the potential difference edge-to-edge between the voltage probes in the nonlocal region. Nonlocal simulations were performed on a 2D $N_r = 61$ by $N_{\theta}=21$ grid, with $\Delta r = 10^{-5}$ and $\Delta \theta = \pi \cdot 10^{-1}$. A time step of $\Delta t = 10^{-15}$ was taken, and 5000 iterations per simulation, each with 100 Poisson steps, were performed.
	\begin{figure}[!hb]
		\centering
		\includegraphics[width=0.75\textwidth]{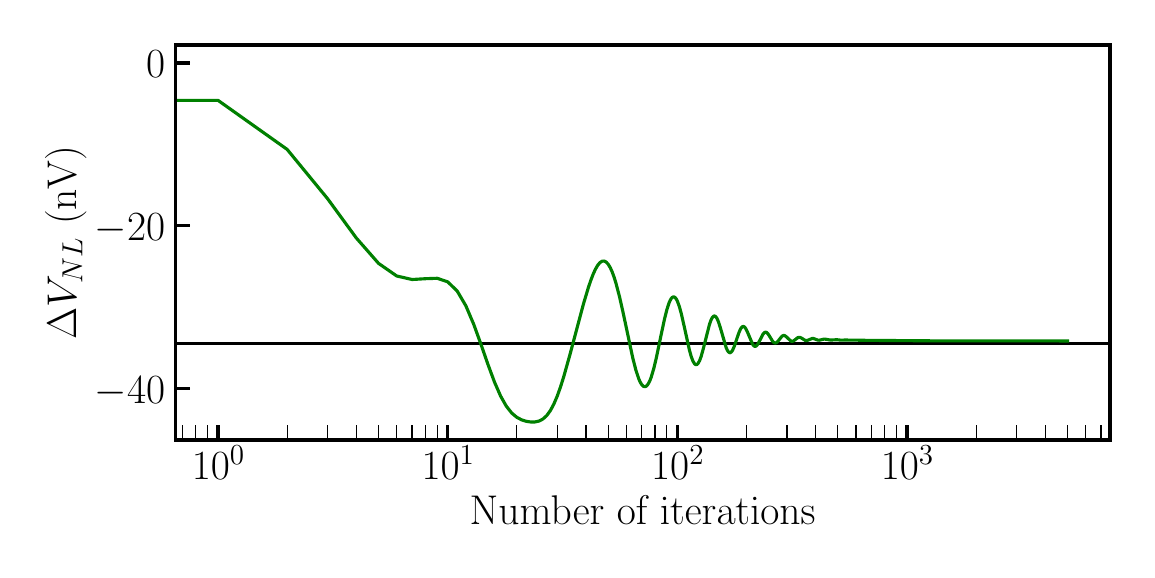}
		\caption{Convergence of nonlocal simulations to analytic solutions. $\Delta V$ for all 5000 time steps of $\Delta t = 10^{-15}~s$ was recorded for CBM302 simulations at $T=1~K$.
			Simulations converge within acceptable limits to the analytic results (shown in black) after 1000 iterations.}
		\label{fig:FigS5}
		
	\end{figure}
	
	\noindent{Due to} the substantial size difference between the inner and outer regions of the Corbino device, the resolution of extrapolated data for reciprocity simulations was limited so severely that we could not derive any meaningful conclusions.\\

	\subsection{AC simulations}
		
	The simulations performed were under the quasi-DC regime, whereby the input current $I_0$ is proportional to the root-mean-square value set on the SR830 lock-in amplifier, $I_{RMS}$, by a factor of $\sqrt{2}$. In this regime, the output would be proportional to the root-mean-square value measured by our lock-in by the same factor.
	
	\noindent{To confirm} our quasi-DC model, we performed AC simulations by simulating uniformly spaced points in our signal period. We then performed a root-mean-square calculation of our AC data points, which yielded {\it nearly} identical results to that of our quasi-DC simulations, shown in \ref{fig:AC}. Each simulation was performed with the results of the previous one in the signal period, adjusted based on
	\begin{figure}[!ht]
		\centering
		\includegraphics[width=0.75\textwidth]{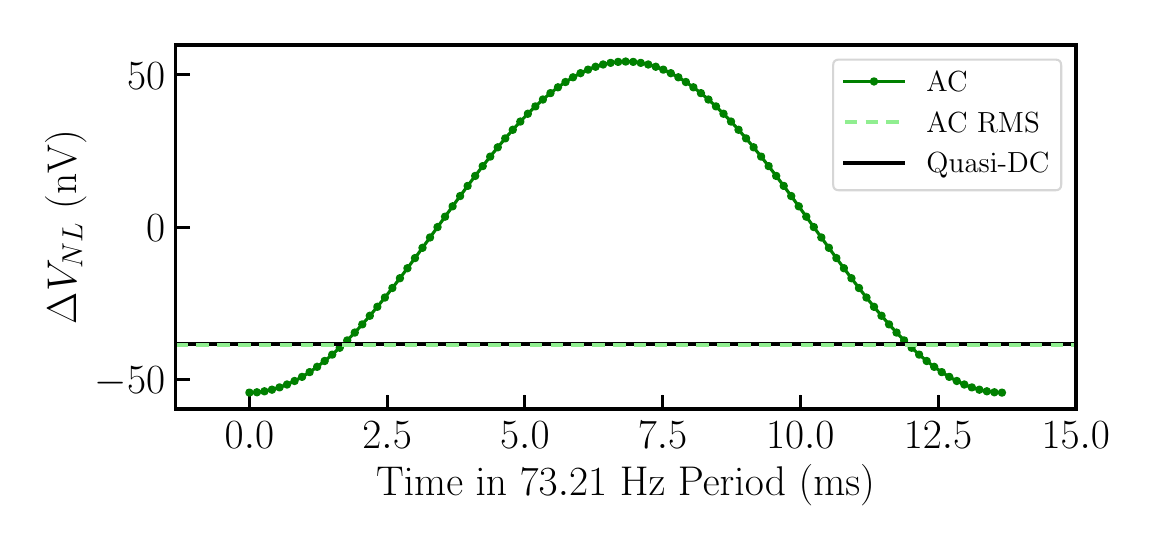}
		\caption{Comparison between AC and quasi-DC. Green points are the resulting $\Delta V$ along one period of a $73.21~\text{Hz}$ AC signal, simulated in CBM302 at base temperature $T=1~K$ for nonlocal configuration. The light green dashed line is the computed root-mean-square value of the AC signal simulated. The black line is the result of the quasi-DC simulation of the same sample and temperature, using the model described in this paper.}
		\label{fig:AC}
	\end{figure}
	
	\begin{eqnarray}
		I(t+\Delta t) &=& I_0 \cos\bigl(\omega (t+\Delta t)\bigr)\nonumber \\
		&=& I_0 \cos(\omega t) \cos(\omega \Delta t) - I_0 \sin(\omega t) \sin(\omega \Delta t)\nonumber \\
		&=& I(t) \Bigl(\cos(\omega \Delta t) - \tan(\omega t)\sin(\omega \Delta t)\Bigr),
	\end{eqnarray}
	where $\omega = 2 \pi f$, and $f$ is frequency.
	\newpage
	\subsection{Validity of extrapolation}
	
	To verify the extrapolation technique used in this study, we performed another set of simulations where we consider the entire annular rings. A key difference between this simulation and the previous one is that we don't assume that velocity $\vec{u} = \int{ \vec{v} f d\vec{v}}$ is equal to the drift velocity $\vec{\lambda}$, defined as $J_{r} = n_e q_e \vec{\lambda}$. As such, starting from the following Boltzmann equation
	\begin{equation}\label{eq:CollisionIntegral}
		\frac{\partial f_{\vec{v}}}{\partial t} + \vec{v}\cdot\frac{\partial f_{\vec{v}}}{\partial \vec{r}} + \frac{q_e}{m^*} \vec{E}\cdot\frac{\partial f^0_{\vec{v}}}{\partial \vec{v}} =
		-\frac{f_{\vec{v}} - f^0_{\vec{v}}}{\tau_{MR}} - \frac{f_{\vec{v}} - f^0_{\vec{\lambda}}}{\tau_{MC}},
	\end{equation}
	the first velocity moment yields the following Navier-Stokes-like equation
	\begin{equation}
		\frac{\partial\vec{u}}{\partial t} +\left(\vec{u} \cdot \vec{\nabla}\right) \vec{u} = \nu \vec{\nabla}^2\vec{u} -\frac{q_e}{m^{*}} \vec{\nabla} \phi - \frac{\vec{u}}{\tau_{MR}} - \frac{\vec{u} - \vec{\lambda}}{\tau_{MC}}.\label{eq:navier_truenonlocal}
	\end{equation}
	
	\noindent{Within} the current path, we set $\vec{u} = \vec{\lambda}$. The potential $\phi = \int{dr \frac{\partial \phi}{\partial r}}$ is calculated directly from Eq.\refeq{eq:navier_truenonlocal} within the current path starting from zero at the grounded inner contact. Outside of the current path, all values are set to zero. The initial conditions are shown in black in Fig. \ref{fig:confirmation}. The boundary conditions of the regions outside the current path region are calculated based on the taylor series centered at the boundary for adjacent values. To determine the derivatives of $\vec{u}$, the constant $K$ from Eq.\refeq{eq:navier2dsoln1} is determined for the adjacent value. For $\phi$, derivatives are calculated directly from Eq.\refeq{eq:navier_truenonlocal} at the boundary.
	
	\noindent{To} prevent errors caused by the discontinuity at the current source, we perform $N_{sub}$ subiterations to determine our guess velocity $\vec{u}^P$ based on the following sequence
	
	\begin{align}
		\vec{u}^{P}_{0} &= \vec{u}^{t}, \\
		\vec{u}^P_{n+1} &= \vec{u}^P_{n} + \frac{\Delta t}{N_{sub}} \left( \nu \vec{\nabla}^2\vec{u}^P_{n} - \left(\vec{u}^P_n \cdot \vec{\nabla}\right) \vec{u}^P_n - \frac{\vec{u}^P_n}{\tau_{MR}} - \frac{\vec{u}^P_n-\vec{\lambda}}{\tau_{MC}}\right),
	\end{align}
	and use $\vec{u}^P_{N_{sub}}$ to solve the Poisson equation. We set $N_{sub} = \lfloor \sqrt{N_r} \rfloor$ as the discontinuity effects scale with $\Delta r \propto \frac{1}{N_r}$.\\
	
	\noindent{The} potential profile and electron velocity, obtained from simulating the entire region over different iterations, along with the result from the extrapolation technique are shown in Fig.\ref{fig:confirmation}. The potential difference between the outer two rings, calculated through the entire region simulation, after 20000 iteration was found to be $-36.5 nV$, and the extrapolation technique gave a value of $-34.1 nV$ after 1458 iterations. . 
	
	\begin{figure}[!htb]
		\centering
		\includegraphics[width=.97\textwidth]{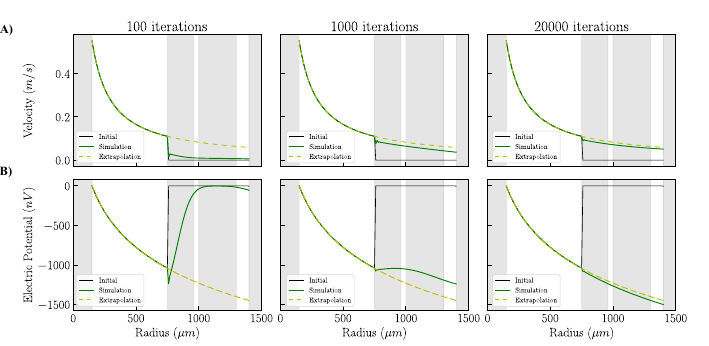}
		\caption{
			Comparison of velocity (A) and electric potential (B) between the extrapolation technique and full region simulation for CBM302 at $T=1~K$ at different number of iterations. The initial conditions of the full region simulation are shown in black, and converges just under 20000 iterations with $N_r=126$, shown in green. The extrapolation technique, depicted by a dashed yellow line, converges after 1458 iterations with $N_r=61$. Both simulations are performed with $\Delta r = 10^{-5}$. The potential profiles from both simulations are in agreement, with minor discrepancies likely due to error propagation from discontinuity at the contact.
		}
		\label{fig:confirmation}
	\end{figure}

\end{document}